\documentstyle[prl,aps,multicol,epsfig]{revtex} 
  
\begin{document} 
\title{Stripe glasses: self generated randomness in a uniformly frustrated
 system} 
\author{J\"org Schmalian$^{(a)}$ and Peter G. Wolynes$^{(b)}$} 
\address{$^{(a)}$ Department of Physics and Astronomy and Ames Laboratory,
Iowa State University, Ames, IA 50011\\ 
$^{(b)}$ Department of Chemistry, University of Illinois at
Urbana-Champaign, Urbana, IL 61801} 
\date{March 15, 2000} 
\maketitle 
\pacs{}
\begin{abstract} 
\leftskip 54.8pt \rightskip 54.8pt We show that a system with competing 
interactions on different length scales, as relevant for the formation of 
stripes in doped Mott insulators, undergoes a self-generated glass 
transition which is caused by the frustrated nature of the interactions and 
not related to the presence of quenched disorder. An exponentially large 
number of metastable configurations is found, leading to a slow, landscape  
dominated long time relaxation 
and a break up of the system into a disordered inhomogeneous state. 
\end{abstract} 
 
\pacs{} 
 
\begin{multicols}{2}  
\narrowtext   
 
Competing interactions on different length scales are able to stabilize 
mesoscale phase separations and the creation of spatial inhomogeneities in a 
wide variety of systems. Examples are stripe formation in doped Mott 
insulators, as found in transition metal oxides (TMO)\cite{CCJ93,JTra95}, 
domains in magnetic multilayer compounds\cite{GD82,AB92}, or \ mesoscopic 
structures formed by assembling polymers in solution and amphiphiles in \ 
water-oil mixtures \cite{GT83,GBS96}. In many of these cases the tendency 
towards a perfectly ordered array of domains, stripes etc. is undermined by 
frustrating long range interactions\cite{EK93}. Very often, these assemblies 
exhibit a long time dynamics similar to the relaxation seen in glasses. In 
the context of stripes it has been argued that the presence of only very few 
quenched impurities might already cause a strictly disordered glassy state  
\cite{KE98}. Furthermore, recent molecular dynamics calculations for charge 
ordering in TMO found an anomalous long time relaxation with a power 
spectrum similar to $1/f$-noise\cite{SYC98}. Indeed, there is experimental 
evidence for the formation of intrinsic inhomogeneities and even a 
stripe-glass in high temperature superconductors and other transition metal 
oxides\cite{CBJ92,LC97,TUI99,JBC99,HS99,CH99,HSS99,MSP00}. In particular 
slow, activated dynamics as observed in NMR experiments \cite{JBC99,CH99} 
exhibits a striking universality, rather independent of the details of added 
impurities etc. It is therefore tempting to speculate that glassiness in 
these systems is {\em self generated }and does not rely on the presence of 
quenched disorder, which may of course further stabilize a glassy state. 
 
In this paper we show that the competition of interactions on different 
length scales in a uniformly frustrated systems exhibits a self generated 
glass transition due to the emergence of an exponentially large number of 
metastable states. This result is obtained using the replica approach of 
Ref.~\cite{Mon95,MP991} and by solving the corresponding many body problem 
using  the self consistent screening
approximation\cite{Bray74,MY92}. Since only very few examples exists 
for models which exhibit self generated glassiness\cite{CIS95,FH95},
all these approaches are extremely important  for a better understanding
of glassiness in general. Even though our findings apply to 
broader class of problems than stripes in TMO, we will adopt a language 
which is specific to that problem.\cite{fn2} 
 
A model for a uniformly frustrated system with competition on different 
length scales is given by the Hamiltonian\cite{EK93}:  
\begin{eqnarray} 
{\cal H} &=&\frac{1}{2}\int d^{3}x\left\{ r_{0}\varphi ({\bf x)}^{2}+\left( 
\nabla \varphi ({\bf x)}\right) ^{2}+\frac{u}{2}\varphi ({\bf x)}^{4}\right\} 
\nonumber \\ 
&&+\frac{Q}{2}\int d^{3}x\int d^{3}x^{\prime }\frac{\varphi ({\bf x)}\varphi 
({\bf x}^{\prime })}{\left| {\bf x-x}^{\prime }\right| }.  \label{ham11} 
\end{eqnarray} 
Here, $\varphi ({\bf x})$ characterizes charge degrees of freedom, with $%
\varphi ({\bf x})>0$ in \ a hole-rich region, $\varphi ({\bf x})<0$ in a 
hole poor region, and $\varphi ({\bf x})=0$ if the local density equals the 
averaged one. If $r_{0}<0$ the system tends to phase separate since we have 
to guarantee charge neutrality, $\left\langle \varphi \right\rangle =0$. The 
coupling constant, $Q$, is a measure for the strength of the Coulomb 
interaction and characterizes the competition between  short 
 and long range interactions. In case of
strongly anisotropic, quasi two-dimensional cuprate superconductors 
one expects an anisotropy of the gradient term in Eq.\ref{ham11}, which we
neglect for simplicity.
Despite the absence of a 
clean derivation of Eq.\ref{ham11} from the many electron Schr\"{o}dinger 
equation, we note that it describes, on a phenomenological level, many of 
the major competing effects which yield in microscopic theories a rich phase 
diagram of inhomogeneous spin and charge structures\cite{VS99}. For $Q=0$ 
and $r_{0}<0$ we expect at low temperatures long range ordered charge 
modulations. As shown in Ref. \cite{NRK99}, the Coulomb interaction 
suppresses this ordered state for all $Q>0$ and finite $T$. Instead, the 
system undergoes several crossovers. Most interestingly, at low 
temperatures, where $\left| r(T)\right| <2\sqrt{Q}$, a mean field analysis 
of Eq.\ref{ham11} shows that besides a correlation length, $\xi =2\left( r+2%
\sqrt{Q}\right) ^{-1/2}$, an additional length scale, $l_{m}=4\pi \left( 2%
\sqrt{Q}-r\right) ^{-1/2}$ emerges, which characterizes the spatial 
modulation of the field correlations\cite{NRK99}, where $r=r_{0}+uT\left%
\langle \varphi ^{2}\right\rangle $.
These modulations are particularly relevant for low 
enough $T$ where $r(T)\leq 0$, where mean field theory gives 
locally ordered regions with characteristic size $l_{m}\ll \xi $.We will 
show that a {\em stripe glass }\ emerges in this temperature regime. 
 
An essential prerequisite for the anomalous dynamical features of 
glassiness, like aging, memory effects and ergodicity breaking, is most 
certainly the occurrence of a large number of metastable states, ${\cal N}%
_{ms}$, separated by energy barriers which are large compared to the 
temperature. In viscous liquids undergoing vitrification calorimetry suggests $%
{\cal N}_{ms}\propto \exp \left( const.V\right) $, were, $V$ is the system 
size. This observation is the heart of an {\em ideal glass transition} 
scenario based on random first order transitions\cite{KTW89} which was 
originally motivated by microscopic stability analyses of structural glasses 
and mean field theories for random Potts-glasses. Below a cross-over 
temperature, $T_{A}$, a 'viscous', energy-landscape dominated long time 
relaxation sets in due to the occurrence of exponentially many metastable 
states, i.e. the configurational entropy, $S_{c}=k_{B}\log {\cal N}_{ms}$, 
becomes extensive. Because of the large barriers between these states, the 
system will get stuck for extremely long times in one of the metastable 
states, i.e. it will freeze into a glass, at some temperature $T_{G}<T_{A}$ 
which depends for example on the cooling rate. Even though this laboratory 
glass transition is purely dynamical, a key ingredient of the ideal glass 
transition scenario is that the dynamical slowing arises from proximity to 
an underlying phase transition at $T_{K}<T_{G}$, where the configurational 
entropy would vanish like $S_{c}(T)\propto T-T_{K}\ $. If such an ideal 
transition exists, even for an infinitely slow cooling rate freezing will 
occur at $T_{K}$ since all the liquid degrees of freedom die out due to this 
'entropy crisis'\cite{AWK48}. 
 
Detailed theoretical investigation of this scenario have concentrated on 
systems with quenched randomness.  A major step forward for studying 
nonrandom systems was made in Ref.~\cite{Mon95,MP991}, where a new replica 
approach was developed. Within this approach, the configurational entropy 
for a model of a structural glass without quenched disorder was calculated 
and found to be in good agreement with computer simulations.\cite{MP991} 
 
We will use this approach to determine $S_{c}$ for a system governed by Eq.%
\ref{ham11}. The key idea is to introduce, in analogy to the theory of 
conventional phase transitions, an appropriate symmetry breaking field, $%
\psi \left( {\bf r}\right) $, and to compute the partition sum  
\begin{equation} 
\widetilde{Z}\left[ \psi \right] =\int D\varphi e^{-{\cal H}[\varphi ]/T-%
\frac{g}{2}\int d^{3}x\left[ \psi \left( {\bf x}\right) -\varphi \left( {\bf %
x}\right) \right] ^{2}},  \label{Zsig} 
\end{equation} 
where $g\rightarrow 0^{+}$. The energy $\widetilde{f}\left[ \psi \right] 
=-T\log \widetilde{Z}\left[ \psi \right] $ will be low if $\psi ({\bf r})$ 
equals to configurations which locally minimize ${\cal H}$. Sampling all 
configurations of the $\psi $-field, weighted with $\exp \left( -\widetilde{f%
}\left[ \psi \right] /T\right) $, is therefore equivalent to scanning all 
metastable states such that  
\begin{equation} 
\widetilde{F}=\lim_{g\rightarrow 0}\frac{1}{W}\int D\psi \text{ }\widetilde{f%
}\left[ \psi \right] \exp \left( -\widetilde{f}\left[ \psi \right] /T\right)  
\label{tild} 
\end{equation} 
is a weighted average of the free energy in the various metastable 
configurations, where $W=$ $\int D\psi \exp \left( -g/2\int d^{3}x\psi 
^{2}\left( {\bf x}\right) \right) $ is introduced for proper normalization. 
If there are only few local minima, the limit $g\rightarrow 0^{+}$ behaves 
perturbatively and $\widetilde{F}$ equals to the free energy, $F$, of the 
system. However, in case exponentially many local minima with large barriers 
between them exist, a nontrivial contributions arises from the $\psi $%
-integral even for $g\rightarrow 0^{+}$ and the averaged free energy, $%
\widetilde{F}$, differs from $F$. This enables us to identify the 
configurational entropy, $S_{c}$, via $F=\widetilde{F}-TS_{c}$.\cite 
{Mon95,MP991} For an illustration of the corresponding free energy 
landscape, see inset of Fig.1. 
 
An explicit expression for $S_{c}$ can be obtained within a replicated 
theory \cite{Mon95} with  
\begin{equation} 
F\left( m\right) =-\lim_{g\rightarrow 0}\frac{T}{m}\log \frac{1}{W}\int D\  
{\bf \psi }\widetilde{Z}^{m}\left[ {\bf \psi }\right] .  \label{fofm} 
\end{equation} 
It follows that $\widetilde{F}=\left. \frac{\partial mF(m)}{\partial m}%
\right| _{m=1}$, which gives:  
\begin{equation} 
S_{c}=\left. \frac{1}{T}\frac{\partial F(m)}{\partial m}\right| _{m=1}. 
\label{conf1} 
\end{equation} 
Inserting $\widetilde{Z}\left[ \psi \right] $ of Eq. \ref{Zsig} into Eq. \ref 
{fofm} finally leads to:  
\begin{eqnarray} 
Z(m) &=&\lim_{g\rightarrow 0}\int D^{m}\varphi \exp \left( -\sum_{a=1}^{m}%
{\cal H}\left[ \varphi ^{a}\right] /T\right.  \nonumber \\ 
&&\text{ \ }\left. -\frac{g}{2m}\sum_{a,b=1}^{m}\int d^{3}x\varphi ^{a}({\bf %
x})\varphi ^{b}({\bf x})\right) ,  \label{repl_part} 
\end{eqnarray} 
with $F\left( m\right) =-\frac{T}{m}\log Z(m)$. Eq.\ref{repl_part} has a 
formal similarity to the action of the random field Ising model, obtained 
within the conventional replica approach, which allows us to use techniques, 
developed for this model\cite{MY92}. In the following we use the self 
consistent screening approximation (SCSA) of Eq.\ref{repl_part} \cite 
{Bray74,MY92,fn3} and determine the Green's function, ${\cal G}_{ab}\left(  
{\bf q}\right) =\left\langle \varphi ^{a}({\bf q)}\varphi ^{b}(-{\bf q)}%
\right\rangle $, in replica space. ${\cal G}_{ab}\left( {\bf q}\right) $ 
then determines the partition function, $Z(m)$, and correspondingly $S_{c}$. 
 
The interaction  between different replicas is 
symmetric with respect to the replica index suggesting the mean field ansatz  
\begin{equation} 
{\cal G}_{ab}\left( {\bf q}\right) =\left( {\cal G}\left( {\bf q}\right) -%
{\cal F}\left( {\bf q}\right) \right) \delta _{ab}+{\cal F}\left( {\bf q}%
\right) ,  \label{repans} 
\end{equation} 
with equal diagonal elements, ${\cal G}\left( {\bf q}\right) $, and equal 
off-diagonal elements, ${\cal F}\left( {\bf q}\right) $\cite{fn1}.The 
physical interpretation of ${\cal G}\left( {\bf r-r}^{\prime }\right) 
=\left\langle \varphi ({\bf r)}\varphi ({\bf r}^{\prime }{\bf )}%
\right\rangle $ as thermodynamic (instantaneous) correlation function is 
straightforward. On the other hand, ${\cal F}\left( {\bf r-r}^{\prime 
}\right) =\lim_{t\rightarrow \infty }\left\langle \varphi ({\bf r,}t{\bf )}%
\varphi ({\bf r}^{\prime },0{\bf )}\right\rangle $ can be interpreted as 
measuring long time correlations, arising from trapping in metastable minima 
which, in mean field theory, have infinite barriers between them. An 
analogous structure in replica space follows for the diagonal elements, $%
\Sigma _{{\cal G}}\left( {\bf q}\right) $, and off-diagonal elements, $%
\Sigma _{{\cal F}}\left( {\bf q}\right) $, of the self energy, which are 
given in the SCSA as:  
\begin{equation} 
\Sigma _{{\cal A}}\left( {\bf q}\right) =2\int \frac{d^{3}p}{\left( 2\pi 
\right) ^{3}}{\cal D}_{{\cal A}}\left( {\bf p}\right) {\cal A}\left( {\bf p+q%
}\right) 
\end{equation} 
with ${\cal A}\in \left\{ {\cal G},{\cal F}\right\} $. The screening of the 
interaction is characterized by the collective propagators ${\cal D}_{{\cal G%
}}^{-1}\left( {\bf p}\right) =(uT)^{-1}+\Pi _{{\cal G}}\left( {\bf p}\right)  
$ and ${\cal D}_{{\cal F}}\left( {\bf p}\right) =\frac{-uT\Pi _{{\cal F}%
}\left( {\bf p}\right) {\cal D}_{{\cal G}}^{2}\left( {\bf p}\right) }{1-uT%
{\cal D}_{{\cal G}}\left( {\bf p}\right) \Pi _{{\cal F}}\left( {\bf p}%
\right) }$ with polarization functions $\Pi _{{\cal A}}\left( {\bf p}\right) 
=\int \frac{d^{3}q}{\left( 2\pi \right) ^{3}}{\cal A}\left( {\bf q+p}\right)  
{\cal A}\left( {\bf q}\right) $. The set of equations is closed by the Dyson 
equation: ${\cal G}^{-1}\left( {\bf k}\right) ={\cal G}_{0}^{-1}\left( {\bf k%
}\right) +\Sigma _{{\cal G}}\left( {\bf k}\right) $ for the diagonal 
elements, and ${\cal F}\left( {\bf k}\right) =\frac{-{\cal G}^{2}\left( {\bf %
k}\right) \Sigma _{F}\left( {\bf k}\right) }{1-{\cal G}\left( {\bf k}\right) 
\Sigma _{F}\left( {\bf k}\right) }$ for the off diagonal elements, 
respectively. ${\cal G}_{0}^{-1}\left( {\bf q}\right) =r+q^{2}+Qq^{-2}$ is 
the inverse Hartree propagator. Note, all momentum integrations have to be 
cut-off at $\left| {\bf p}\right| =\Lambda $, which is of the order of an 
inverse lattice constant. Once the ${\cal G}_{ab}$ and ${\cal D}_{ab}$ are 
determined the free energy becomes  
\begin{equation} 
F(m)/(2mT)=\text{tr}\log {\cal G}^{-1}+\text{tr}\log {\cal D}^{-1}-\text{tr}%
\Sigma {\cal G}\,. 
\end{equation} 
After performing the trace in replica space for arbitrary integer $m$ and 
analytical continuation to $m\rightarrow 1$, $S_{c}$ follows from Eq.\ref 
{conf1}. One finds immeadiately $S_{c}=0$ if  ${\cal F}({\bf k})$ vanishes.
In the following we discuss the numerical solution of this set of coupled
 integral equations. 
 
\begin{figure}
\centerline{\epsfig{file=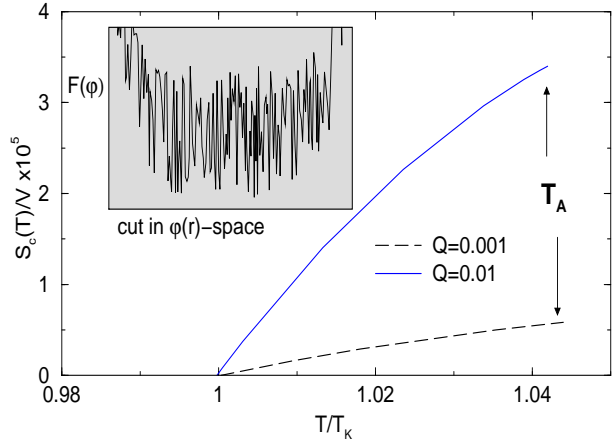,width=8cm,height=6cm,scale=0.85}} 
\caption{Configurational entropy density, $S_{c}/V$ as function of $T/T_{K}$ 
for $Q=0.01$ and $r_{0}=-10$ ($T_{K}=1.7586$) and $Q=0.001$ and $r_{0}=-6$ ($%
T_{K}=1.0422$). \ In both cases $u=\Lambda =1$ is used. Note the strong 
$Q$-dependece of $S_{c}$.
The inset shows a typical energy landscape for finite $S_{c}$.} 
\label{fi2} 
\end{figure} 
In Fig. 1, $S_{c}/V$ is shown for two different $Q$-values as 
function of $T$. At $T_{A}$
the long time correlation function ${\cal F}\left( k\right) $ emerges 
leading to $S_c>0$ and a glassy dynamics sets in. 
The corresponding free energy landscape is schematically illustrated in the 
inset. Just as in mean field Potts glasses with quenched 
randomness\cite{KTW89}, $S_{c}$ vanishes  at a lower
 temperature, $T_{K}$. At $T_{K} 
$ the entropy of the amorphous stripe solid equals that of the stripe 
liquid. There is no entropic advantage anymore to be in a liquid state, 
leading to an obligatory glass transition no matter how slow the cooling 
rate. The laboratory glass temperature, $T_{G}$, will lie somewhere between $%
T_{K}$ and $T_{A}$ and cannot be determined within our theory. We 
also find that $T_{K}$ and $T_{A}$ are only weakly 
decreasing for increasing $Q$, see inset of 
Fig.2. Both temperatures remain finite for $Q\rightarrow 0$. However, $%
S_{c}\left( Q\rightarrow 0\right) \rightarrow 0$, i.e. the fragility $%
\propto \frac{dS_{c}}{dT}$ of the glass vanishes.
 In other words, the larger the modulation length, the smaller 
is the number of metastable states. Due to the ${1}/{r}$ 
interaction, the limit $Q\rightarrow 0$  does not smoothly 
connect to the behavior at $Q=0$. If one includes a finite screening 
length $l_{s}$ in Eq.\ref{ham11}, we expect that the glassy state disappears 
for $Q \ll l_{s}^{-4}$. 
\begin{figure}
\centerline{\epsfig{file=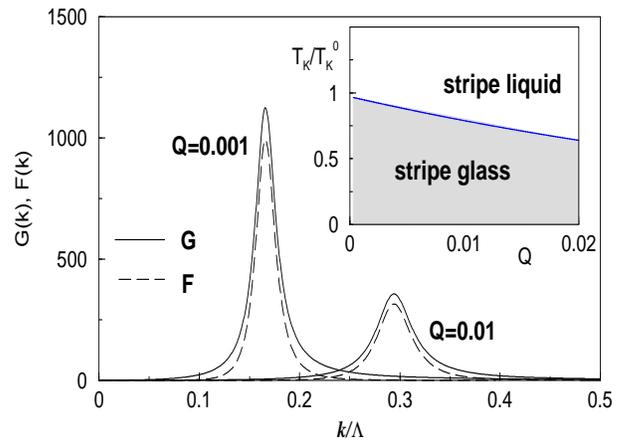,width=8cm,height=6cm,scale=0.85}} 
\caption{Momentum dependence of the instantaneous (${\cal G}$) and long-time 
(${\cal F}$) charge correlation function for the same parameters as in Fig 1 
at $T=T_{A}$. The inset shows the glass liquid phase diagram as function of $%
Q$ for $r_{0}=-10$.} 
\label{fi1} 
\end{figure} 
 
In Fig. 2, the instantaneous (${\cal G}\left( k\right) $) and long-time ($%
{\cal F}\left( k\right) $) charge correlation functions, are shown at $%
T=T_{A}$. Even though no charge ordering occurs, the pronounced peaks at 
finite $k$ demonstrate that there is a modulated state with strong short 
range correlations, $l_{m} < \xi $. Also, since ${\cal F}\left( k\right) 
\lesssim {\cal G}\left( k\right) $, the modulated state exhibits an 
anomalous dynamics, where long time correlations are only slightly 
reduced compared to instantaneous correlations. Introducing a Debye-Waller 
factor, $W=-\log \left( {\cal F}/{\cal G}\right) $ for $k$ close to the peak 
maximum, gives $W=0.12$ $(0.13)$ for $Q=0.01$ $(0.001)$. The  
modulation length  (inverse peak positions) is $3.5 \Lambda^{-1}$ 
($6 \Lambda^{-1}$) and the correlation length is $45 \Lambda^{-1}$
($80 \Lambda^{-1}$) for $Q=0.01$ $(0.001)$.

Due to the competing interactions in Eq.\ref{ham11}, an 
entropy crisis occurs, causing a transition into a glass. This purely 
thermodynamic characterization of the spectrum of metastable states, is only 
the first important step for understanding glassiness, and the investigation 
of dynamical features is an even bigger challenge, because it requires going 
beyond mean field theory. An argument based on ''entropic droplets'' 
explains quite well the phenomenology of viscous liquid dynamics\cite{KTW89} 
and can even be made semi-quantitative\cite{XW99}. Here we apply these 
arguments to the present stripe model. The entropic droplet argument 
recognizes an intrinsic instability of the homogeneous  
metastable solutions, as characterized by Eq.~\ref{repans}: 
namely, creating a droplet of one metastable solution within another costs 
free energy that can at most scale as a surface energy but the exponential 
number of configurations gives an entropic driving force for such a droplet 
that scales with V. A mosaic structure hence will form. The activation 
free energy of turning over a single region can be computed
 where the entropic gain is given by  $TS_{c}$.  A 
renormalization group calculation, based on Ref.\cite{Villain84}, leads to a 
size dependent surface tension $\sigma (R)=\sigma _{0}\left( R\Lambda 
\right) ^{-\theta }$ with $\theta =\frac{d-2}{2}$ reflecting the fact that 
the interface between two states is wetted by intermediate states. This 
analysis leads to an characteristic energy barrier $\Delta E\propto \left( 
TS_{c}(T)\right) ^{-1}$ which implies a  relaxation time 
obeying a Vogel-Fulcher law\cite{KTW89}  
\begin{equation} 
\tau \propto \exp \left( \frac{DT_K}{T-T_{K}}\right) . 
\end{equation} 
An estimate for the surface tension $\sigma _{0}\approx |r_0|/(u\xi)$
for the stripe model yields 
$R_0^3\approx \Lambda^{-1}(V\sigma_0/TS_c)^2$ for the droplet volume and
$D=3V\sigma_0^2/(\Lambda T^{2}T_K\left. \partial S_{c}/\partial T\right| 
_{T_{K}})$. Using our numerical results for $S_c$, this leads to 
$D\approx 60 - 200$, typical for moderately fragile and strong glasses,
 and droplet sizes
$R_0\approx (25-50)\Lambda^{-1}\approx (5-10)l_m$. Note that this estimate is
only qualitative since $R_0 \approx \xi$  and a real
separation of scales never occurs. 
The droplet picture implies that the glass state breaks up into domains of 
different metastable states, separated by wetted surfaces, build by 
intermediate states. This physical 
picture is very similar to the conclusions made in Ref.\cite{CBJ92} based on 
NMR experiments. 
 
In summary, we have shown that an exponentially large number of metastable 
configurations emerges in a system with competing interactions on
different length scales, leading to a glass transition and anomalous long time 
dynamics. This glass state is self generated, implying that the barriers 
characterizing the activated dynamics are rather universal and should not 
depend on details like added impurities but only on the generic interactions 
on short and long scales, i.e. the magnetic exchange interactions and the 
Coulomb interaction. Furthermore, we showed that the magnitude of the 
frustration controls the fragility of the glass transition.
Finally, following Ref.\cite{KTW89}, we argued 
that the configurational entropy causes a break up of the stripe glass into 
a mosaic of domains or droplets, build up by the various metastable
 states, allowing us to estimate time scales of motions. 
This causes an intrinsic inhomogeneity of all relevant correlation 
functions, modulation and correlation lengths etc. in the amorphous glassy 
state. 
 
We gratefully acknowledge stimulating discussions at the workshop on 
{\em Mesoscopic organization in soft hard and biological matter} hosted by
the Institute for Complex Adaptive Matter and with A. V. Chubukov, P. C. 
Hammel, J. Haase, D. C. Johnston, D. K. Morr, D. Pines, C. P. Slichter, R. 
Stern and B. P. Stojkovic. The work was supported by NSF (PGW), grant No.ChE-9530680.  Ames Laboratory is operated for the U.S. 
Department of Energy by Iowa State University under Contract No. 
W-7405-Eng-82.

\end{multicols}  
 

\begin{references}
\bibitem{fn2}  Eq.\ref{ham11} is a continuum model and will not give one 
dimensional stripes but rather spherically symmetric ring pattern as 
modulated structures. However, as shown in Ref.\cite{NRK99}, the inclusion 
of\ small lattice corrections yields stripe type pattern. We will use the 
term 'stripe' in this loose sense.
\bibitem{fn3}  The applicability of this approximation is supported by the 
non-singular behavior of the equivalent large-$N$ limit of Eq.\ref{ham11}, 
as shown in Ref.\cite{NRK99}.
\end{references}
\end{document}